\documentclass[aps,prd,twocolumn,showpacs,showkeys,superscriptaddress]{revtex4-1}
\usepackage{latexsym}
\usepackage{amssymb}
\usepackage{amsmath}
\usepackage{amscd}
\usepackage{amsthm}
\usepackage{graphicx}
\usepackage{textcomp}
\usepackage{colortbl}
\usepackage{hyperref}
\usepackage[font={footnotesize,it}]{caption}
\usepackage{multirow}
\usepackage[T1]{fontenc}
\usepackage{ae,aecompl}

\begin{document}

\title{Comment on Brans-Dicke Scalar Field Cosmological Model in Lyra's Geometry}
\author{Anil Kumar Yadav}
\email{abanilyadav@yahoo.co.in}
\affiliation{Department of Physics, United College of Engineering and Research,Greater Noida - 201310, India}

\begin{abstract}
In the present study, we comment on Brans-Dicke scalar field cosmological model in Lyra's geometry [Maurya \& Zia, Phys. Rev. D \textbf{100}, 023503 (2019)]. In this comment, we investigate that there is no acceleration in the model proposed by the authors of Phys. Rev. D \textbf{100}, 023503 (2019). Therefore, despite the claims to the contrary the Brans-Dicke Scalar Field Cosmological Model in Lyra's Geometry with high Brans-Dicke (BD) coupling parameter $\omega$ and constant $\beta$ can not produce late time acceleration in the universe. 
\end{abstract}



\maketitle
\section{Introduction}
\label{sec:intro}
Today, General theory of Relativity (GR) is credited the most successful theory to describe the late time acceleration with inclusion of some type of exotic/dark energy in Einstein's field equation. In 1915, Einstein was formulated GR and tried to give a satisfactory description to Mach's principle. But soon, he realized that his theory does not follow the Mach's principle. In the last century, some theories of gravitation had been postulated to follow Mach's principle. In 1961, Brans and Dicke \cite{Brans/1961} had proposed a scalar-tensor relativistic theory of gravitation in which Mach's principle is validated. With validation of Mach's principle, the proposed Brans-Dicke theory also describes the  inflation era but it requires extra dark matter candidates to explain the galactic velocity profiles\cite{Das/2012}. The concept of geometrizing gravitation in the form of GR gives a clue to researchers to think about the geometrizing electromagnetic field also. Weyl \cite{Weyl/1918} had proposed a geometrized theory of electromagnetism and gravitation which is based the on non-integrability of length transfer. In 1951, Lyra \cite{Lyra/1951} had proposed geometrized theory without non-integrability condition. In subsequent investigations, several authors \cite{Sen/1957,Sen/1971,Halford/1970,Halford/1972} have constructed cosmological model in the framework of Lyra's geometry. It is shown from these investigations that the scalar-tensor treatment based on Lyra's geometry predicts the same effects, as GR, under observational limits.   
\section{Theoretical model and Basic equations}
\label{sec:2}
The Bianchi I space-time is read as \cite{Maurya/2019}
\begin{equation}
ds^{2} = -c^{2}dt^{2}+A^{2}dx^{2}+B^{2}dy^{2}+C^{2}dz^{2}
\end{equation}
where A, B \& C are directional scale factors and they are functions of $t$ only. In Ref.\cite{Maurya/2019}, $c^{2}$ is missed from equation (16) but appeared in subsequent equations. \\
The Einstein's Brans-Dicke field equations in Lyra's manifold is read \cite{Maurya/2019} as
\[
G_{ij}+\frac{3}{2}\psi_{i}\psi_{j}-\frac{3}{4}g_{ij}\psi_{k}\psi^{k} =
\]
\begin{equation}
\label{lm}
 -\frac{8\pi T_{ij}}{\phi c^{4}} -\frac{\omega}{\phi^{2}}\left(\phi_{,i}\phi_{,j}-\frac{1}{2}g_{ij}\phi_{,k}\phi^{,k}\right) -\frac{1}{\phi}(\phi_{,i,j}-g_{ij}\square \phi)
\end{equation}
\begin{equation}
\label{lm-1}
\square \phi = \phi^{,i}_{,i} = \frac{8\pi T}{(3+2\omega)c^{2}}
\end{equation}
where $G_{ij}$, $\psi^{i}$, $\omega$ and $\phi$ are Einstein's curvature tensor, displacement vector field of Lyra's geometry, Brans-Dicke coupling constant and scalar field respectively. Also the time like constant displacement vector is read as $\psi_{i} = (\beta,0,0,0)$. In equation (\ref{lm}), $T_{ij}$ denotes the energy-momentum tensor of perfect fluid.\\ 

As given in Ref. \cite{Maurya/2019}, the gravitational field equations for Bianchi I space-time are read as
\begin{equation}
\label{fe-1}
\frac{\dot{A}\dot{B}}{AB}+\frac{\dot{B}\dot{C}}{BC}+\frac{\dot{A}\dot{C}}{AC}-\frac{3}{4}\beta^{2}=\frac{8\pi \rho}{\phi c^{2}}-\frac{\omega}{2}\frac{\dot{\phi}^{2}}{\phi^{2}}+\frac{\dot{\phi}}{\phi}\left(\frac{\dot{A}}{A}+\frac{\dot{B}}{B}+\frac{\dot{C}}{C}\right)
\end{equation}
\begin{equation}
\label{fe-2}
\frac{\ddot{B}}{B}+\frac{\ddot{C}}{C}+\frac{\dot{B}\dot{C}}{BC}+\frac{3}{4}\beta^{2}=-\frac{8\pi p}{\phi c^{2}}+\frac{\omega}{2}\frac{\dot{\phi}^{2}}{\phi^{2}}+\frac{\dot{\phi}}{\phi}\left(\frac{\dot{B}}{B}+\frac{\dot{C}}{C}\right)+\frac{\ddot{\phi}}{\phi}
\end{equation}
\begin{equation}
\label{fe-3}
\frac{\ddot{A}}{A}+\frac{\ddot{C}}{C}+\frac{\dot{A}\dot{C}}{AC}+\frac{3}{4}\beta^{2}=-\frac{8\pi p}{\phi c^{2}}+\frac{\omega}{2}\frac{\dot{\phi}^{2}}{\phi^{2}}+\frac{\dot{\phi}}{\phi}\left(\frac{\dot{A}}{A}+\frac{\dot{C}}{C}\right)+\frac{\ddot{\phi}}{\phi}
\end{equation}
\begin{equation}
\label{fe-4}
\frac{\ddot{A}}{A}+\frac{\ddot{B}}{B}+\frac{\dot{A}\dot{B}}{AB}+\frac{3}{4}\beta^{2}=-\frac{8\pi p}{\phi c^{2}}+\frac{\omega}{2}\frac{\dot{\phi}^{2}}{\phi^{2}}+\frac{\dot{\phi}}{\phi}\left(\frac{\dot{A}}{A}+\frac{\dot{B}}{B}\right)+\frac{\ddot{\phi}}{\phi}
\end{equation}
\begin{equation}
\label{fe-5}
\frac{\ddot{\phi}}{\phi}+\frac{\dot{\phi}}{\phi}\left(\frac{\dot{A}}{A}+\frac{\dot{B}}{B}+\frac{\dot{C}}{C}\right) = \frac{8\pi(\rho-3p)}{(2\omega+3)\phi c^{2}}
\end{equation}
Note that equations (\ref{fe-1})-(\ref{fe-5}) are same as the equations (17) - (20) in Maurya and Zia \cite{Maurya/2019}. It is worth to noting that the field equations (17) - (20) in Maurya and Zia \cite{Maurya/2019} are wrong (see Refs. \cite{Brans/1961},\cite{Faraoni/2004},\cite{Goswami/2017},\cite{Sen/2001},\cite{Narlikar/2002}). The correct field equations are as follows.
\begin{equation}
\label{fe-1c}
\frac{\dot{A}\dot{B}}{AB}+\frac{\dot{B}\dot{C}}{BC}+\frac{\dot{A}\dot{C}}{AC}-\frac{3}{4}\beta^{2}=\frac{8\pi \rho}{\phi c^{2}}+\frac{\omega}{2}\frac{\dot{\phi}^{2}}{\phi^{2}}-\frac{\dot{\phi}}{\phi}\left(\frac{\dot{A}}{A}+\frac{\dot{B}}{B}+\frac{\dot{C}}{C}\right)
\end{equation}
\begin{equation}
\label{fe-2c}
\frac{\ddot{B}}{B}+\frac{\ddot{C}}{C}+\frac{\dot{B}\dot{C}}{BC}+\frac{3}{4}\beta^{2}=-\frac{8\pi p}{\phi c^{2}}-\frac{\omega}{2}\frac{\dot{\phi}^{2}}{\phi^{2}}-\frac{\dot{\phi}}{\phi}\left(\frac{\dot{B}}{B}+\frac{\dot{C}}{C}\right)-\frac{\ddot{\phi}}{\phi}
\end{equation}
\begin{equation}
\label{fe-3c}
\frac{\ddot{A}}{A}+\frac{\ddot{C}}{C}+\frac{\dot{A}\dot{C}}{AC}+\frac{3}{4}\beta^{2}=-\frac{8\pi p}{\phi c^{2}}-\frac{\omega}{2}\frac{\dot{\phi}^{2}}{\phi^{2}}-\frac{\dot{\phi}}{\phi}\left(\frac{\dot{A}}{A}+\frac{\dot{C}}{C}\right)-\frac{\ddot{\phi}}{\phi}
\end{equation}
\begin{equation}
\label{fe-4c}
\frac{\ddot{A}}{A}+\frac{\ddot{B}}{B}+\frac{\dot{A}\dot{B}}{AB}+\frac{3}{4}\beta^{2}=-\frac{8\pi p}{\phi c^{2}}-\frac{\omega}{2}\frac{\dot{\phi}^{2}}{\phi^{2}}-\frac{\dot{\phi}}{\phi}\left(\frac{\dot{A}}{A}+\frac{\dot{B}}{B}\right)-\frac{\ddot{\phi}}{\phi}
\end{equation}
Accordingly equations (27) - (30) are not correct in Refs. \cite{Maurya/2019}. The procedure of solving the above equations are described below:\\
Subtracting  equations (\ref{fe-2c}) from (\ref{fe-3c}), (\ref{fe-3c}) from (\ref{fe-4c}) and (\ref{fe-4c}) from (\ref{fe-2c}), we obtain the following system of equations 
\begin{equation}
\label{ef-7}
\frac{\ddot{A}}{A}-\frac{\ddot{B}}{B}+\frac{\dot{A}\dot{C}}{AC}-\frac{\dot{B}\dot{C}}{BC}+\left(\frac{\dot{A}}{A}-\frac{\dot{B}}{B}\right)\frac{\dot{\phi}}{\phi} = 0
\end{equation}
\begin{equation}
\label{ef-8}
\frac{\ddot{B}}{B}-\frac{\ddot{C}}{C}+\frac{\dot{A}\dot{B}}{AB}-\frac{\dot{A}\dot{C}}{AC}+\left(\frac{\dot{B}}{B}-\frac{\dot{C}}{C}\right)\frac{\dot{\phi}}{\phi} = 0
\end{equation}
\begin{equation}
\label{ef-9}
\frac{\ddot{C}}{C}-\frac{\ddot{A}}{A}+\frac{\dot{B}\dot{C}}{BC}-\frac{\dot{A}\dot{B}}{AB}+\left(\frac{\dot{C}}{C}-\frac{\dot{A}}{A}\right)\frac{\dot{\phi}}{\phi} = 0
\end{equation}
The equations (\ref{ef-7})-(\ref{ef-9}) are not closed because this is the system of three equations with four unknown variables $A$, $B$, $C$ and $\phi$. So, the solutions of these equations are not possible in general. However, to obtain the explicit solution of above equations, one has to assume a viable relation among variables. For this sake, we consider the following relation among the directional scale factors as
\begin{equation}
\label{s-1}
B = AD \;\;\; \& \;\; C = \frac{A}{D}
\end{equation}
where $D = D(t)$ denotes anisotropy in the universe of derived model.\\

Equations (\ref{ef-8}) and (\ref{s-1}) lead to
\begin{equation}
\label{s-2}
\frac{\ddot{D}}{D}-\frac{\dot{D}^{2}}{D^{2}}+\frac{\dot{D}}{D}\left(3\frac{\dot{A}}{A}+\frac{\dot{\phi}}{\phi}\right) = 0
\end{equation}
After integration of equation (\ref{s-2}), we obtain
\begin{equation}
\label{s-3}
D = exp\left[\int\frac{k}{A^{3}\phi}dt\right]
\end{equation} 
Now, the average scale factor is computed as 
\begin{equation}
\label{s-2}
a^{3} = ABC = A^{3} \Rightarrow a = A
\end{equation}
In light of equation (\ref{s-1}), equations (\ref{fe-1c}) and (\ref{fe-2c}) take the following form
\begin{equation}
\label{fe-6}
3\frac{\dot{A}^{2}}{A^{2}}-\frac{\dot{D}^{2}}{D^{2}} -\frac{3}{4}\beta^{2} = \frac{8\pi \rho}{\phi c^{2}}+\frac{\omega}{2}\frac{\dot{\phi}^{2}}{\phi^{2}}-3\frac{\dot{\phi}}{\phi}\frac{\dot{A}}{A} 
\end{equation}
\begin{equation}
\label{fe-7}
2\frac{\ddot{A}}{A}+\frac{\dot{A}^{2}}{A^{2}}+\frac{\dot{D}^{2}}{D^{2}} +\frac{3}{4}\beta^{2} = -\frac{8\pi p}{\phi c^{2}}-\frac{\omega}{2}\frac{\dot{\phi}^{2}}{\phi^{2}}-2\frac{\dot{\phi}}{\phi}\frac{\dot{A}}{A}-\frac{\ddot{\phi}}{\phi} 
\end{equation}
where $\frac{\dot{D}}{D} = \frac{k}{A^{3}\phi}$ with $k$ as the arbitrary constant.\\

The deceleration parameter $q$ and Huble's parameter $H$ are defined as
\begin{equation}
\label{dp}
q = -\frac{a\ddot{a}}{\dot{a}^{2}} = -\frac{A\ddot{A}}{\dot{A}^{2}}
\end{equation}
\begin{equation}
\label{h-1}
H= \frac{\dot{a}}{a} = \frac{\dot{A}}{A}
\end{equation}
Differentiating equation (\ref{h-1}), we obtain
\begin{equation}
\label{h-2}
\dot{H} = \frac{\ddot{A}}{A} -\frac{\dot{A}^{2}}{A^{2}} = \frac{\ddot{A}}{A} - H^{2}
\end{equation}
Now, deceleration parameter in terms of H is given by
\begin{equation}
\label{dp-2}
q = -1-\frac{\dot{H}}{H^{2}}
\end{equation}
Also, we define the matter energy density parameter $\Omega_{m}$, anisotropy parameter $\Omega_{\sigma}$, $\beta$ parameter $\Omega_{\beta}$ and $q_{\phi}$ as \cite{Maurya/2019}
\begin{equation}
\label{ed}
\Omega_{m} = \frac{8\pi\rho}{3c^{2}H^{2}\phi},\;\;\Omega_{\sigma} = \frac{k^{2}}{3H^{2}A^{6}\phi^{2}},\;\;\Omega_{\beta}=\frac{\beta^{2}}{4H^{2}},\;\;q_{\phi}=-\frac{\ddot{\phi}}{\phi H^{2}}
\end{equation}
Using equations (\ref{h-1})-(\ref{ed}) and after some algebra, equations (\ref{fe-6}), (\ref{fe-7}) and (\ref{fe-5}) lead to
\begin{equation}
\label{fe-8}
\Omega_{m}+\Omega_{\sigma}+\Omega_{\beta} = 1-\frac{\omega}{6}\Psi^{2}+\Psi
\end{equation}
\begin{equation}
\label{fe-9}
2q=1+3(\gamma\Omega_{m}+\Omega_{\beta}+\Omega_{\sigma})+\frac{\omega}{2}\Psi^{2}+2\Psi-q_{\phi}
\end{equation}
\begin{equation}
\label{fe-10}
-q_{\phi}+3\Psi = \frac{3(1-3\gamma)}{2\omega+3}\Omega_{m}
\end{equation}
where $\Psi = \frac{\dot{\phi}}{\phi H}$.\\
Solving equations (\ref{fe-8}), (\ref{fe-9}) and (\ref{fe-10}), we obtain 
\begin{equation}
\label{dp-3}
q=2+\Psi-\frac{3[(1-\gamma)\omega+1]}{2\omega+3}\Omega_{m}
\end{equation} 
Also, it is easy to find that the main equations of the model in standard BD cosmology by introducing two effective parameters as
$$\rho_{eff} = \rho+\frac{3\phi c^{2}}{32 \pi}\beta^{2}$$ \& $$p_{eff} = p+\frac{3\phi c^{2}}{32 \pi}\beta^{2}$$ 
Thus, equations (\ref{fe-6}) and (\ref{fe-7}) are recast as
\begin{equation}
\label{eff-1}
3\frac{\dot{A}^{2}}{A^{2}}-\frac{\dot{D}^{2}}{D^{2}} = \frac{8\pi}{\phi c^{2}}\rho_{eff}+\frac{\omega}{2}\left(\frac{\dot{\phi}}{\phi}\right)^{2}-3\frac{\dot{A}\dot{\phi}}{A\phi}
\end{equation}
\begin{equation}
\label{eff-2}
2\frac{\ddot{A}}{A}+\frac{\dot{A}^{2}}{A^{2}}+\frac{\dot{D}^{2}}{D^{2}} = -\frac{8\pi}{\phi c^{2}}p_{eff}-\frac{\omega\dot{\phi}^{2}}{2\phi^{2}}-2\frac{\dot{A}\dot{\phi}}{A\phi}-\frac{\ddot{\phi}}{\phi}
\end{equation} 
For analysis of model, it is convenient to consider the effective equation of state parameter $(\omega_{eff})$ as
$$\omega_{eff} = \frac{p_{eff}}{\rho_{eff}} = \frac{p+\frac{3\phi c^{2}}{32 \pi}\beta^{2}}{\rho+\frac{3\phi c^{2}}{32 \pi}\beta^{2}}$$
Thus, we observe that in absence of matter, the effective equation of state parameter is equal to $+1$ $i. e.$ for $ \rho = p =0$, we obtain $\omega_{eff} = +1$. That is why the displacement vector can not play the role of a cosmological constant in Brans-Dicke theory for which $\omega_{eff} = -1$ is required. It is worth to noting that in Ref. \cite{Maurya/2019}, $\omega_{eff} \neq -1$.\\
\section{Discussion}
It is well known that for accelerating cosmological model $q < 0$. In the derived model, the fluid under consideration is perfect fluid therefor $p \geq 0$. Also  $\Psi=\frac{\dot{\phi}}{\phi H}$ is positive.\\

Thus, from equation (\ref{dp-3}), for accelerating cosmological model, we have
\[
2+\Psi-\frac{3[(1-\gamma)\omega+1]}{2\omega+3}\Omega_{m} < 0
\]
\begin{equation}
\label{dp-4}
\Rightarrow 2+\Psi < \frac{3[(1-\gamma)\omega+1]}{2\omega+3}\Omega_{m}
\end{equation}
\begin{figure}[h!]
\includegraphics[width=8cm,height=6cm,angle=0]{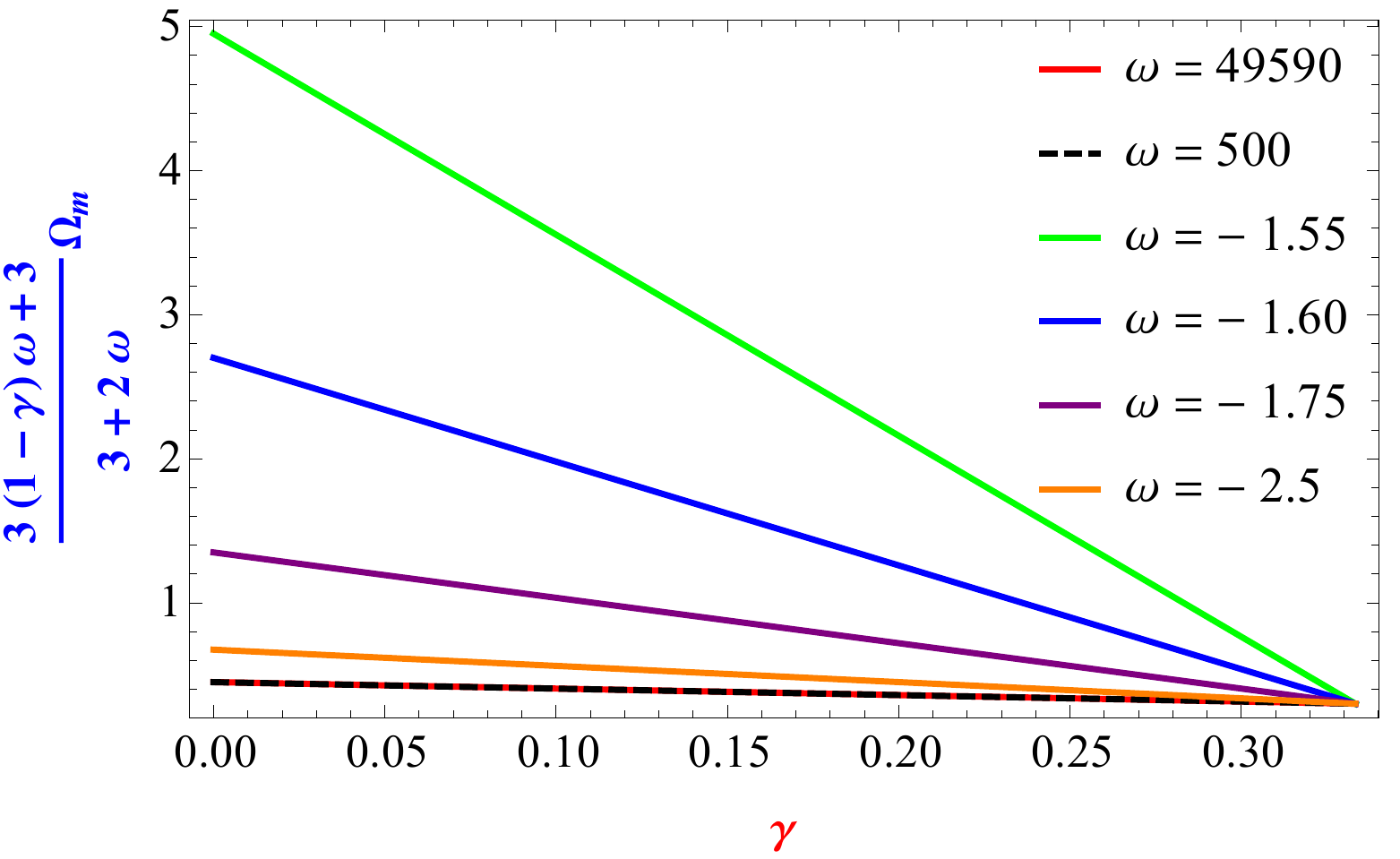}
\caption{The plot of $\frac{3(1-\gamma)(\omega+1)}{2\omega+3}$ versus $\gamma$ for different numerical values of $\omega$.}
\label{fig1}
\end{figure}
From equation (\ref{dp-4}), we conclude that in the derived model acceleration is only possible when $\frac{3[(1-\gamma)\omega+1]}{2\omega+3}\Omega_{m} > 2$ because $2 + \Psi$ is always greater than 2. The plot of numerically computed values of  $\frac{3[(1-\gamma)\omega+1]}{2\omega+3}\Omega_{m}$ versus $0 \leq \gamma < 1/3$ for some particular values of $\omega$ is shown in Fig. 1. We choose $\gamma$ in the range $ 0 \leq \gamma < 1/3$ for numerical result and analysis of model because the authors of Ref. \cite{Maurya/2019} have taken this range of $\gamma$ in describing the late acceleration of the universe. We observe that for $\omega = 49590$, the derived model does not validate equation (\ref{dp-4}) because the value of $\frac{3[(1-\gamma)\omega+1]}{2\omega+3}\Omega_{m}$ is less than 2 (see Fig. 1). From numerical result plotted in Fig. 1, we also observe that for small negative values of BD coupling parameter (i. e. $\omega$ = - 1.55 and $\omega$ = -1.60) the late time acceleration is possible in the derived model. Therfore, in spite of mathematical errors in Maurya and Zia \cite{Maurya/2019}, the late time acceleration is not possible with large BD coupling parameter ($\omega = 49590$). This result is in favor of investigations presented in Refs. \cite{Sen/2001,Banerjee/2001}. In Akarsu et al. \cite{Akarsu/2020}, some specific negative ranges of $\omega$ are given which describe acceleration in massive Brans-Dicke gravity.\\ 
\begin{figure}[h!]
\includegraphics[width=8cm,height=6cm,angle=0]{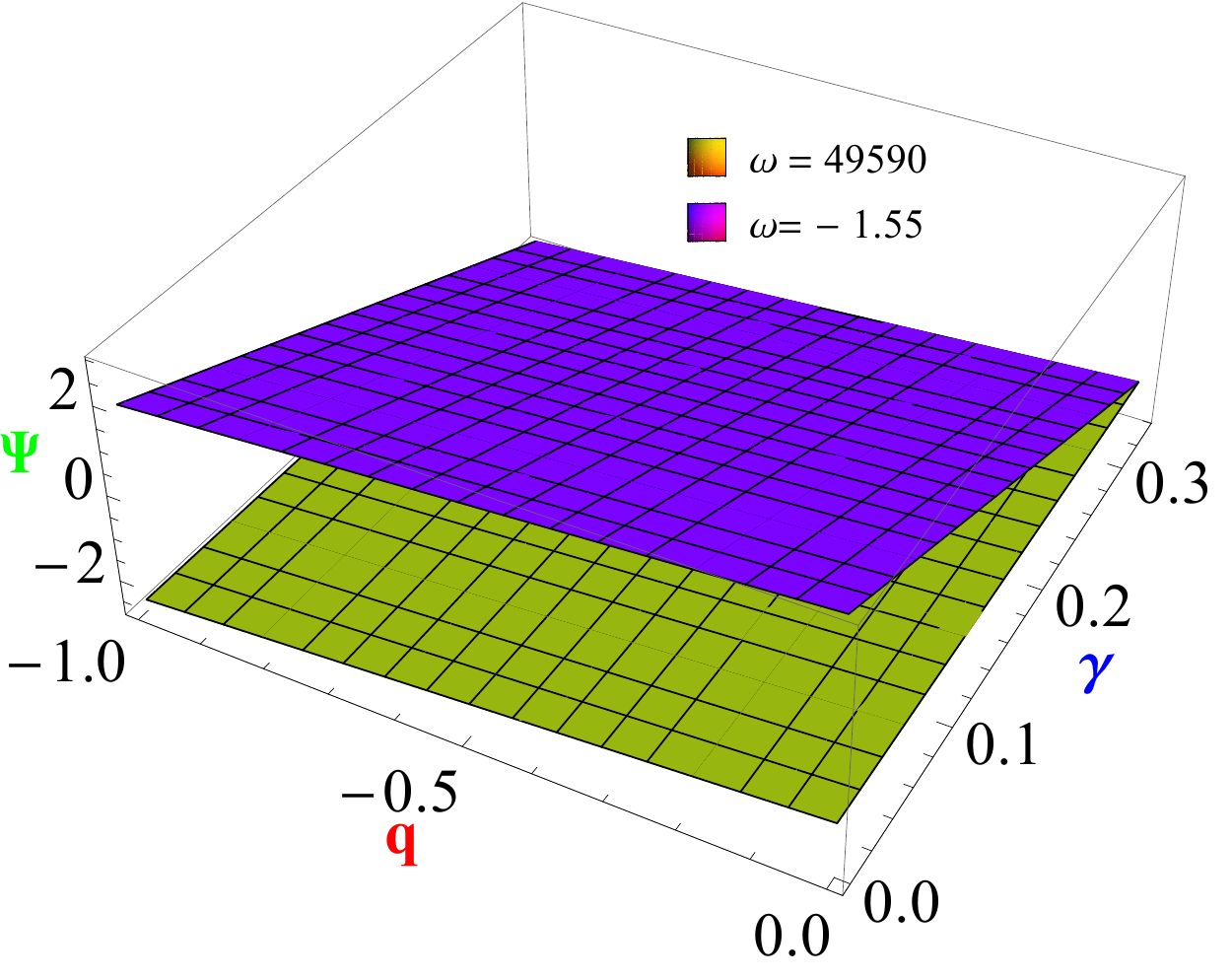}
\caption{The plot of $\Psi$ versus $\gamma$ and $q$ for $\omega = 49590$ and $\omega = - 1.55$.}
\label{fig2}
\end{figure}
The variation of $\Psi$ versus $q$ and $0 \leq \gamma < 1/3$ for $\omega = 49590$ and $\omega = -1.55$ is depicted in Fig. 2. Since $\dot{\phi} > 0$ therefore $\Psi = \frac{\dot{\phi}}{\phi H}$ is always greater than zero in Brans-Dicke theory of gravitation (see Ref. \cite{Faraoni/2004}). From Fig. 2, we observe that for $\omega = 49590$, in Maurya and Zia \cite{Maurya/2019}, positive $\Psi$ and negative $q$ do not exist simultaneously while $\omega = - 1.55$ validates this condition.\\
 
Therefore, in the derived model, the late time acceleration is produced due to amalgamation of BD theory with small negative BD coupling parameter (adverse of author's finding in Ref. \cite{Maurya/2019}). One of the important fearture of Brans-Dicke theory is that BD theory of gravity is able to fulfill the requirement of Mach's principle\cite{Brans/1961,Fujii/2003,Faraoni/2004,Uehara/1982,Lorenz/1984}. In Sen and Sen \cite{Sen/2001}, it has been argued that the late time acceleration in the universe is driven by dissipative pressure. Also, the summary of investigation of Ref. \cite{Sen/2001} give a clue that the late time acceleration without inclusion of $\Lambda$ or quintessence matter in the energy/matter of the universe, in Brans-Dicke cosmology, is possible for small and negative Brans-Dicke coupling constant \cite{Banerjee/2001} while in Ref. \cite{Bertolami/2000}, it has been investigated that $\phi^{2}$ potential for large Brans-Dicke coupling constant without considering positive energy condition for matter and scalar field both may describe the late time acceleration in the universe. In the targeted paper\cite{Maurya/2019}, the authors have not clearly argued that the late time acceleration is due to scalar field. They focused on gauge function of Lyra's geometry and hypothetically assumed that $\beta = constant$ behaves like cosmological constant $\Lambda$. Further they argued that it may be a suitable candidate of dark energy and removes the cosmological constant problems while the investigations in Lyra's geometry clearly established the fact that time varying displacement vector $\beta(t)$ have the similar nature as $\Lambda(t)$ $i.  e.$ $\beta(t)$ and $\Lambda(t)$ both are decreasing function of time \cite{Singh/1997}. It is important to note that this decreasing behaviour of $\beta(t)$ with time does not contibute late time acceleration in the universe \cite{Yadav/2018}.  But in Maurya and Zia \cite{Maurya/2019}, it is conveniently assumed that constant $\beta$ play the role of cosmological constant and leads the late time acceleration of the universe without giving concrete mathematical expression or exact physical reason behind it. \\

The continuity equation in Lyra's geometry is read as
\begin{equation}
\label{ce}
\chi\dot{\rho}+\frac{3}{2}\beta\dot{\beta}+3\left[\chi(\rho+p)+\frac{3}{2}\beta^{2}\right]H = 0
\end{equation}  
where $\chi = \frac{8\pi}{c^{4}}$.\\
The equation (22) of targeted paper is entirely different from equation (\ref{ce}) of this comment. It seems that authors have assumed only General Relativity case in energy conservation law. \\

For $\beta = constant$, equation (\ref{ce})is given by
\begin{equation}
\label{ce-1}
\chi\dot{\rho}+3\left[\chi(\rho+p)+\frac{3}{2}\beta^{2}\right]H = 0
\end{equation}
Thus, the energy conservation law given in Ref.\cite{Maurya/2019} is not correct in context of Lyra's geometry which turns into imply that the matter energy density as given in equation (53) of targeted paper, may have different expression. In addition, we observe that for constant displacement vector $i. e$ $\beta = constant$, there is a constant contribution to equations (\ref{fe-1})- (\ref{fe-4}). Therefore, despite the claims to the contrary made by the authors, the model cannot be consistent with observations. Some important applications of Lyra's geometry with time varying displacement vector are given in the Refs. \cite{Yadav/2018,Beesham/1988,Singh/1991,Singh/1993,Beesham/1993,Yadav/2011}.
\section{Conclusion}
\label{sec:3}
In this comment, we have shown that the field equations derived in Ref. \cite{Maurya/2019}, are not contributing late time acceleration with constant $\beta$ and large $\omega$ but late time acceleration in the model may be the feature of universe due to Brans-Dicke scalar field that have small negative value of $\omega$. However, the actual physics of such acceleration with large BD coupling constant is not elaborated in Maurya and Zia \cite{Maurya/2019}. It is convenient to assume $\beta = $ constant but this constant displacement vector does not contribute the late time acceleration of the universe with large $\omega$. We have also corrected the field equations and subsequent equations which were wrong in Maurya and Zia \cite{Maurya/2019}. It is worth to note that neither we avoid the co-existence of BD scalar field with Lyra's geometry nor decline the similarities between time varying displacement vector $\beta(t)$ and $\Lambda(t)$ as both $\Lambda(t)$ and $\beta(t)$ are decreasing function of time. As a final comment, we note that in spite of good possibility of scalar field cosmological model in Lyra's geometry to provide a theoretical foundation for relativistic cosmology, the experimental point is yet to be considered and still the theory needs a fair trial.    

\section*{ACKNOWLEDGMENT}
The author is grateful to the editor and referee for illuminating suggestions that have significantly improved our work in terms of research quality and presentation. We also acknowledge G. K. Goswami for careful reading of manuscript prior to submission.

\end{document}